\documentclass[12pt]{article}
\usepackage{amssymb} \usepackage{amsmath}  \usepackage{graphicx}

\begin{document}
\title{Instabilities in neutrino-plasma density waves}
\author{Lu\'{\i}s Bento  \\  
%\address{
{\em Centro de F\'{\i}sica Nuclear da Universidade de Lisboa,} \\
{\em Av. Prof. Gama Pinto 2, 1649-003 Lisboa, Portugal}     } 
\date{December, 2000}
\maketitle

\begin{abstract}
One examines the interaction and possible resonances between supernova
neutrinos and electron plasma waves. The neutrino phase space distribution
and its boundary regions are analyzed in detail. It is shown that the
boundary regions are too wide to produce non-linear resonant effects. The
growth or damping rates induced by neutrinos are always proportional to the
neutrino flux and $G_{{\rm F}}^{2}$.
\end{abstract}

\strut


%\baselineskip=15pt

%\pacs{PACS numbers: 95.30.Qd, 13.15.+g, 14.60.Lm, 97.60.Bw}

\newpage
\baselineskip=15pt

\section{Introduction}

The study of the interactions between neutrinos and plasma waves has
received a great deal of attention. Early works of Bingham {\em et~al.}~\cite
{bingham94,dawson98}\ claimed that an intense neutrino flux passing through
a plasma is capable of producing unstable modes of electron density waves
causing a significant transfer of energy from neutrinos to the mantle of a
Supernova. If true, this could constitute a sort of realization of the
Wilson explosion mechanism. However, their description of the weak
interactions in the neutrino-electron system was not satisfactory.

More recently~\cite{bento98,bento99}, the full Standard Model Quantum Field
Theory of electroweak interactions was applied to establish the dynamics of
the excitations of the electromagnetic field and electron and neutrino
current density distributions (Eqs.~(27)-(32) of~\cite{bento99}). From that
we derived the modification on the dispersion relation of electron density
waves due to a neutrino flow. Our analysis of the conditions of neutrino
emission in Supernovae led us to the conclusion that they do not satisfy the
necessary requirements to generate the unstable waves and growing rates
predicted by Bingham {\em et~al.~}\cite{bingham94,dawson98}. Subsequent
works~\cite{nieves00,elze00} gave some support to this conclusion. However,
the controversy seems to persist on this specific issue~\cite{silva99}. We
report here a detailed analysis concerning the neutrino phase space
distribution and implications for plasma waves.

\section{Neutrino induced instabilities}

The dispersion relation of electron density waves, also called plasmons, is
modified by weak interactions when a neutrino flow passes through the
plasma. Let $\omega _{pl}({\bf k})$\ designate the wave frequency as a
function of the wavevector ${\bf k}$ in absence of neutrinos and in the
plasma collisionless limit. We assume the background medium to be static and 
{\em spatially} homogeneous within the time and length scales characteristic
of the plasma waves of interest. The momentum distribution is assumed
isotropic for electrons but not for neutrinos, as they have an almost unique
direction at far distances from the Supernova core. The electron plasma is
considered non-relativistic. In these conditions a stream of neutrinos
(anti-neutrinos) modifies the dispersion relation as follows~\cite
{bento98,bento99} ($\hbar =c=1$) 
\begin{equation}
\omega ^{2}-\omega _{pl}^{2}=\Gamma A\,\omega _{pl}^{2}\;,  \label{dr}
\end{equation}
\begin{equation}
\Gamma =2G_{{\rm F}}^{2}\,c_{V}^{\prime 2}\,\frac{n_{e}n_{\nu }}{m_{e}\bar{E}%
_{\nu }}\;,  \label{gama}
\end{equation}
where $c_{V}^{\prime }=1/2+2\sin ^{2}\theta _{{\rm W}}\simeq 0.96$ for $\nu
_{e}$ ($\bar{\nu}_{e}$), $G_{{\rm F}}$ is the Fermi constant, $n_{e}$, $%
n_{\nu }$ are the electron and neutrino number densities, respectively, and 
\begin{equation}
A=\left( 1-\frac{\omega ^{2}}{{\bf k}^{2}}\right) ^{2}\frac{{\bf k}^{2}\bar{E%
}_{\nu }}{\omega _{p}^{2} \, n_{\nu }} \int \!d^{3}p_{\nu }\frac{f_{\nu }}{%
E_{\nu }}\frac{{\bf k}^{2}\!-({\bf k}\!\cdot \!{\bf v}_{\nu })^{2}}{(\omega -%
{\bf k}\!\cdot \!{\bf v}_{\nu })^{2}-(\omega ^{2}-{\bf k}^{2})^{2}/4E_{\nu
}^{2}}\;.  \label{Aquantic}
\end{equation}
is a dimensionless quantity. $f_{\nu }$\ is the neutrino or anti-neutrino\
distribution function in momentum space, $\int \!f_{\nu }d^{3}p_{\nu
}=n_{\nu }$, and $\bar{E}_{\nu }$ is a typical neutrino energy. Finally, $%
\omega _{p}^{2}=4\pi \alpha n_{e}/m_{e}$.

The expression $A$ admits in general a classical approximation because the
frequency and wavenumber are much smaller than the single particle energy $%
E_{\nu }$. In fact $\omega $ is around the magnitude of the plasma
frequency, $\omega _{p}$, and $k$ is limited above by the Debye wavenumber, $%
k_{{\rm D}}=\omega _{p}\sqrt{m_{e}/T_{e}}$ at a temperature $T_{e}~$\cite
{lifshitz97}. In the classic limit $A$ is approximated as 
\begin{equation}
A=\left( 1-\frac{\omega ^{2}}{{\bf k}^{2}}\right) ^{2}\frac{{\bf k}^{2}%
\bar{E}_{\nu }}{\omega _{p}^{2}\,n_{\nu }}\int \!d^{3}p_{\nu }\frac{f_{\nu }%
}{E_{\nu }}\frac{{\bf k}^{2}-({\bf k}\!\cdot \!{\bf v}_{\nu })^{2}}{(\omega -%
{\bf k}\!\cdot \!{\bf v}_{\nu })^{2}}\;,  \label{Ahydrodynamic}
\end{equation}
or, after integrating by parts, 
\begin{equation}
A=-\left( 1-\frac{\omega ^{2}}{{\bf k}^{2}}\right) ^{2}\frac{{\bf k}^{2}%
\bar{E}_{\nu }}{\omega _{p}^{2}\,n_{\nu }}\int \!d^{3}p_{\nu }\frac{{\bf %
k\cdot }\partial f_{\nu }/\partial {\bf p}_{\nu }}{\omega -{\bf k}\!\cdot \!%
{\bf v}_{\nu }}\;,  \label{Akinetic}
\end{equation}
an expression directly related to classic kinetic theory~\cite
{bento98,bento99}.

The frequency shift due to weak interactions is in general extremely small.
The factor $G_{{\rm F}}^{2}n_{e}n_{\nu }/m_{e}\bar{E}_{\nu }$ is about $3\times
10^{-28}$ for electron and neutrino densities as high as $n_{e}=10^{29}\,%
{\rm cm}^{-3}$ and $n_{\nu }=L_{\nu }/4\pi r^{2}\bar{E}_{\nu }=1.8\times 10^{30}\,%
{\rm cm}^{-3}$ at radius $r=300\,{\rm km}$ for a neutrino energy luminosity $%
L_{\nu }=10^{52}\,{\rm erg/s}$ and $\bar{E}_{\nu }=10\,{\rm MeV}$. The claim
has been~\cite{bingham94,dawson98,silva99} that the shift on the imaginary
part of the frequency, $\gamma ={\rm Im}\{\omega -\omega _{pl}\}$, is not
suppressed by $G_{{\rm F}}^{2}$\ but rather by a smaller power of $G_{{\rm F}%
}$\ for certain wave modes that are resonantly enhanced by powers of $%
(\omega -{\bf k}\!\cdot \!{\bf v}_{\nu })^{-1}\sim \gamma ^{-1}\gg \omega
_{pl}^{-1}$. That would be the case if all the neutrinos had exactly the
same velocity vector, ${\bf v}_{\nu }$. Then Eq.~(\ref{Ahydrodynamic}) would
give $A\sim \omega _{pl}^{2}/\gamma ^{2}$\ for the resonant modes and the
dispersion relation (\ref{dr}) complex solutions $\omega ({\bf k})$ with
growth rates $\gamma \sim \Gamma ^{1/3}\omega _{pl}$,\ around $10^{10}{\rm s}
$ for the parameters shown above. This corresponds to the reactive
instability put forward by Bingham {\em et~al.}~\cite{bingham94,dawson98}.

However, that calculation is not realistic because the neutrinos do not have
exactly the same direction of motion. As emphasized in~ \cite
{bento98,bento99}, no matter how far the neutrinos are from the core, there
is always an angular spread proportional to the neutrinosphere radius, $R$, $%
2\alpha _{\nu }\simeq 2R/r$. This causes a variation of $\omega -{\bf k}%
\!\cdot \!{\bf v}_{\nu }$ proportional to $\alpha _{\nu }\omega _{pl}$,
orders of magnitude higher than any conceivable value for the width $|\gamma
|$ of a resonance due to neutrino interactions. This means that $\omega
_{pl}-{\bf k}\!\cdot \!{\bf v}_{\nu }$ changes sign over the neutrino
momentum distribution and only a vanishing small fraction of the neutrinos
lie on the resonance. Such a situation is similar to Landau damping \cite
{lifshitz97,melrose86} in an electron plasma and the imaginary part of $A$
can be calculated by replacing $(\omega -{\bf k}\!\cdot \!{\bf v}_{\nu
})^{-1}$ with $-i\pi \,\delta (\omega _{pl}-{\bf k}\!\cdot \!{\bf v}_{\nu })$%
. It yields a neutrino contribution to $\gamma $ proportional to $G_{{\rm F}%
}^{2}$ given by Eq.~(\ref{gamaLandau}), first obtained by Hardy and Melrose~ 
\cite{hardy97} through other methods. The only possible exceptions, that we
want to analyze here, are resonances located at some boundary of the phase
space occupied by neutrinos.

One kinematic boundary is the velocity direction parallel to ${\bf k}$ (when 
${\bf k}$\ lies inside the neutrino velocity cone): it corresponds to the
minimum angle between ${\bf v}_{\nu }$ and ${\bf k}$ and minimum value of $%
\omega _{pl}-{\bf k}\!\cdot \!{\bf v}_{\nu }$. The other boundary is the
upper limit on the angle between ${\bf v}_{\nu }$\ and the radial direction, 
$\theta \leq \alpha _{\nu }\simeq R/r$,\ due to the finite size of the
neutrinosphere. Ideally, the distribution function would be discontinuous at 
$\theta =\alpha _{\nu }$\ but that is not true as we will see.

Consider first the case of a resonance at ${\bf v}_{\nu }$\ parallel to $%
{\bf k}$ i.e., $\omega _{pl}=k\,v_{\nu }$. Then, the factor $(1-\omega ^{2}/%
{\bf k}^{2})^{2}$ in $A$ vanishes unless the neutrinos are massive in which
case $v_{\nu }\simeq 1-m_{\nu }^{2}/2E_{\nu }^{2}$. Let $v_{0}=\omega
_{pl}/k $ be the exact resonant speed. Then $(1-\omega ^{2}/{\bf k}%
^{2})\simeq m_{\nu }^{2}/E_{0}^{2}$ and $\omega _{pl}-{\bf k}\!\cdot \!{\bf v%
}_{0}$\ is positive throughout the neutrino angular distribution. The
problem is, the very mass that makes $1-\omega ^{2}/{\bf k}^{2}$ different
from zero also causes a neutrino speed variation over the energy spectrum
and a change of sign in $\omega _{pl}-{\bf k}\!\cdot \!{\bf v}_{\nu }$. In
fact, $\omega _{pl}-kv_{\nu }$ is negative for energies larger than $E_{0}$.
Only energies obeying $\omega _{pl}-kv_{\nu }\lesssim |\gamma |$ can
participate in the resonance of width $|\gamma |$. The standard deviation $%
\Delta E_{\nu }$ in the energy spectrum of Supernova neutrinos~\cite
{mayle87,burrows90} is comparable to the average energy $\bar{E}_{\nu }$, $%
\Delta E_{\nu }\approx \bar{E}_{\nu }/2.5$, which implies a deviation $%
\Delta v_{\nu }\approx 2m_{\nu }^{2}/5\bar{E}_{\nu }^{2}$. In order that a
significant fraction of the energy spectrum contributes to the resonance it
is necessary that $\Delta v_{\nu }\lesssim |\gamma |/\omega _{pl}$, but that
puts an upper limit on the neutrino mass, $m_{\nu }^{2}/\bar{E}_{\nu
}^{2}\lesssim 5|\gamma |/\omega _{pl}$, and, quite remarkably, on the factor 
\begin{equation}
1-\omega ^{2}/{\bf k}^{2}\simeq \frac{m_{\nu }^{2}}{E_{0}^{2}}\lesssim \frac{%
5|\gamma |}{\omega _{pl}}\;.  \label{factor limit}
\end{equation}
That simply washes out the resonance because the numerator $(1-\omega ^{2}/%
{\bf k}^{2})^{2}$ in $A$ becomes suppressed by $\gamma ^{2}$. To put it in
another way, if there is a resonance for ${\bf v}_{\nu }$\ parallel to ${\bf %
k}$, the energy spectrum is always too broad to prevent $\omega
_{pl}-kv_{\nu }$\ from changing sign and departing from the resonance width.

The other phase space boundary is the upper limit on the angle $\theta $
between ${\bf v}_{\nu }$\ and the radial direction, $\alpha _{\nu }\simeq R/r
$ ($R$ is the neutrinosphere radius and $r$ the distance from the Supernova
center). If the resonance lies on $\theta =\alpha _{\nu }$, i.e., $\omega
_{pl}=k\cos \alpha _{\nu }$, then $\omega _{pl}-{\bf k}\!\cdot \!{\bf v}%
_{\nu }$ is essentially negative over the neutrino angular distribution.
However, the distribution does not fall abruptly to zero at the polar angle $%
\alpha _{\nu }$. There are two reasons for this. First, the neutrinosphere
radius depends on the $\nu $ energy: the interaction cross sections increase
with the $\nu $ energy and consequently more energetic neutrinos suffer the
last scattering in regions of lower density, farther from the center.
Second, scattering is a statistical process by nature. Particles with the
same energy suffer the last scattering at different radii according to a
certain statistical distribution dependent upon the particular chemical and
density profile of the medium (see Appendix). Both factors imply that the
neutrinosphere has a considerable thickness and so has the neutrino angular
aperture. This fact changes the way the distribution function depends on $%
\theta $.

Let $R_{E}$ and $\Delta R_{E}$ be the average neutrinosphere radius and
statistical uncertainty for neutrinos with well defined energy $E_{\nu }=E$
and $\alpha _{E}=R_{E}/r$, $\Delta \alpha _{E}=\Delta R_{E}/r$ the
respective angular aperture and uncertainty. Assuming axial symmetry around
the radial direction, the distribution function only depends on the energy
and polar angle: $f_{\nu }=f_{\nu }(E,\theta )$. Its derivative with respect
to $\theta $ can be modeled as 
\begin{equation}
\frac{\partial f_{\nu }}{\partial \theta }(E,\theta )=-\frac{g(E)}{\sqrt{%
2\pi }\Delta \alpha _{E}}e^{-%
%TCIMACRO{\tfrac{(\theta -\alpha _{E})^{2}}{2\Delta \alpha _{E}^{2}}}%
%BeginExpansion
{\textstyle{(\theta -\alpha _{E})^{2} \over 2\Delta \alpha _{E}^{2}}}%
%EndExpansion
}\;,  \label{fderivative}
\end{equation}
with $g(E)\simeq f_{\nu }(E,0)$. It corresponds to a distribution function
practically constant in the interval $\theta <\alpha _{E}-2\Delta \alpha
_{E} $ and dropping to zero at $\theta >\alpha _{E}+2\Delta \alpha _{E}$. In
addition to $\Delta \alpha _{E}$, the distribution function is also smoothed
by the dependence of $\alpha _{E}$\ on the neutrino energy, which makes a
total angular width $\Delta \alpha _{\nu }=\Delta \alpha _{\bar{E}}+{\rm d}%
\alpha _{E}/{\rm d}E_{\nu }\,\Delta E_{\nu }$, centered on the polar angle $%
\alpha _{\nu }$. The consequence of this is that $\omega _{pl}-{\bf k}%
\!\cdot \!{\bf v}_{\nu }$\ may change sign and depart from the resonance
still within the angular boundary $\theta \approx \alpha _{\nu }\pm 2\Delta
\alpha _{\nu }$,\ if the resonance is not wide enough.

Let ${\bf v}_{0}$\ be a particular vector of the resonance surface (defined
by ${\bf k}\!\cdot \!{\bf v}_{\nu }=\omega _{pl}$) situated in the angular
boundary i.e., with polar angle $\theta _{0}$ close to $\alpha _{\nu }$, and
azimutal angle $\phi _{0}.$ The angular coordinates of a generic velocity
vector ${\bf v}_{\nu }$ and wavevector ${\bf k}$ are denoted as $(\theta
,\phi )$ and $(\theta _{k},\phi _{k})$, respectively. Without loss of
generality, $\phi _{k}=0$. The product ${\bf k}\!\cdot \!{\bf v}_{\nu }$\
and its variation from ${\bf v}_{0}$\ are given by 
\begin{eqnarray}
{\bf k}\!\cdot \!{\bf v}_{\nu } &=&kv_{\nu }(\cos \theta _{k}\cos \theta
+\sin \theta _{k}\sin \theta \cos \phi )\;,  \label{kv} \\
\delta {\bf k}\!\cdot \!{\bf v}_{\nu } &=&k_{\theta }v_{0}\,\delta \theta
+k_{\phi }v_{0}\sin \theta _{0}\,\delta \phi \;,  \label{dkv}
\end{eqnarray}
where $k_{\theta }=k(-\cos \theta _{k}\sin \theta _{0}+\sin \theta _{k}\cos
\theta _{0}\cos \phi _{0})$\ and $k_{\phi }=-k\sin \theta _{k}\sin \phi _{0}$%
\ represent the \ components of ${\bf k}$ along the directions ${\bf e}%
_{\theta }$ and ${\bf e}_{\phi }$, respectively. With $k_{v}={\bf k}\!\cdot
\!{\bf v}_{0}/v_{0}$\ they obey the identity $k_{\theta }^{2}+$ $k_{\phi
}^{2}+k_{v}^{2}=1$. The angular displacement 
\begin{eqnarray}
\delta \theta  &=&\pm \frac{\gamma \,\cos \beta }{\sqrt{{\bf k}%
^{2}v_{0}^{2}-({\bf k}\!\cdot \!{\bf v}_{0})^{2}}}\;,  \label{dteta} \\
\sin \theta _{0}\,\delta \phi  &=&\pm \frac{\gamma \,\sin \beta }{\sqrt{{\bf %
k}^{2}v_{0}^{2}-({\bf k}\!\cdot \!{\bf v}_{0})^{2}}}\;,  \label{dfi}
\end{eqnarray}
with $\tan \beta =k_{\phi }/k_{\theta }$, causes a variation $\delta {\bf k}%
\!\cdot \!{\bf v}_{\nu }=\pm \gamma $. Notice that ${\bf k}\!\cdot \!{\bf v}%
_{0}=\omega _{pl}$ and $\sin \theta _{0}\simeq \sin \alpha _{\nu }$. If that
displacement satisfies $|\delta \theta |\ll \Delta \alpha _{\nu }$\ and $%
|\delta \phi |\ll \pi $, then $\omega _{pl}-{\bf k}\!\cdot \!{\bf v}_{\nu }$%
\ goes away to both sides of the resonance for velocity directions well
inside the range $\theta \approx \alpha _{\nu }\pm 2\Delta \alpha _{\nu }$
and therefore well inside the neutrino distribution. Of course, $\omega
_{pl}-{\bf k}\!\cdot \!{\bf v}_{\nu }$\ would have a definite sign if $%
\theta _{0}>\alpha _{\nu }\pm 2\Delta \alpha _{\nu }$ but then, the
resonance would exist where the neutrino distribution function drops to zero
i.e., there are no neutrinos.

Our estimates of the neutrinosphere depth (see Appendix) give $\Delta \alpha
_{\nu }$ varying between $0.11\,\alpha _{\nu }$ and $0.03\,\alpha _{\nu }$
at different stages of Supernova evolution and $\alpha _{\nu }=R/r$ between $%
0.1$ and $0.03$ at $r=300$ km. On the other hand, $|\gamma |$ is orders of
magnitude below $10^{-9}\omega _{pl}$. This implies that the displacements
calculated above satisfy 
\begin{eqnarray}
|\delta \theta |/\Delta \alpha _{\nu } &\lesssim &10^{3}\,|\gamma |/(\omega
_{pl}\sin \theta _{k\nu })\;,  \label{dtetadalfa} \\
|\delta \phi |/\pi  &\lesssim &10\,|\gamma |/(\omega _{pl}\sin \theta _{k\nu
})\;,  \label{dfipi}
\end{eqnarray}
and therefore are well inside the neutrino angular boundary. The contrary
would require a very small angle between ${\bf k}$ and ${\bf v}_{0}$, $%
\theta _{k\nu }<10^{-6}$ for $|\gamma |/\omega _{pl}<$\ $10^{-9}$, which
leads to a huge suppression factor $(1-\omega ^{2}/{\bf k}^{2})^{2}\simeq
\theta _{k\nu }^{4}$ in $A$, which in turn further suppresses the values of $%
\gamma $ and so on, in other words, it falls in the case ${\bf v}_{0}$
parallel to ${\bf k}$ treated above in first place.

The lesson from all this is that the contribution, $\gamma _{\nu }$, of
neutrino weak interactions to the imaginary part of the wave frequency $%
\omega ({\bf k})$, is too small to produce resonances where the quantity $A$
is inversely proportional to some power of $\gamma _{\nu }$. On the
contrary, $A$ is independent of $\gamma _{\nu }$\ and 
\begin{equation}
\gamma _{\nu }=\frac{1}{2}\Gamma \,\omega _{pl}\,{\rm Im}\{A\}\;.
\label{gamaniu}
\end{equation}
Applying Landau prescription~\cite{lifshitz97,melrose86} to Eq.~(\ref
{Akinetic}), the result is 
\begin{equation}
\frac{\gamma _{_{{\rm \nu }}}}{\omega _{pl}}=\frac{G_{{\rm F}%
}^{2}\,c_{V}^{\prime \,2}}{4\alpha \,{\bf k}^{2}}({\bf k}^{2}-\omega
_{pl}^{2})^{2}\hspace{-0.3044pc}\int \hspace{-0.3033pc}d^{3}p_{\nu }\,\delta
(\omega _{pl}-{\bf k}\!\cdot \!{\bf v}_{\nu })\,{\bf k\cdot }\frac{\partial
f_{\nu }}{\partial {\bf p}_{\nu }}\ ,  \label{gamaLandau}
\end{equation}
the same as obtained by Hardy and Melrose~\cite{hardy97} from the study of
stimulated emission and absorption of plasmons by neutrinos. $\gamma _{\nu }$%
\ is proportional to $G_{{\rm F}}^{2}$ and suppressed by $G_{{\rm F}%
}^{2}n_{e}n_{\nu }/m_{e}\bar{E}_{\nu }$, down to $\sim 10^{-28}$ even for
electron and neutrino densities as high as $n_{e}=10^{29}\,{\rm cm}^{-3}$
and $n_{\nu }\sim 10^{30}\,{\rm cm}^{-3}$ at $300\,{\rm km}$ from the
Supernova center. The other point is, the rate $\gamma _{\nu }$\ does not
drive the evolution of plasma waves because it is many orders of magnitude
smaller than the damping rate caused by electron-ion collisions, $\gamma _{c}
$, only two or three orders of magnitude below $\omega _{pl}$ for such high
density plasmas~ \cite{lifshitz97,melrose86}.

More important than the growth rate is to know the energy transferred from
neutrinos to the plasma. Keeping in mind that the overall wave growth is
shut down by electron-ion collisional damping, the energy transferred per
unity of time to a single mode is $2\gamma _{\nu }({\bf k})\omega _{pl}({\bf %
k})$. Assuming that the plasma waves obey a thermal equilibrium
Bose-Einstein distribution, cut-off at the Debye wavenumber $k_{{\rm D}}$,
the total energy transferred per unity of time and volume is 
\begin{equation}
\dot{\rho}=2\int \!\frac{d^{3}k}{(2\pi )^{3}}\frac{\gamma _{\nu }({\bf k}%
)\omega _{pl}}{e^{\omega _{pl}/T_{e}}-1}\;.  \label{rhodot}
\end{equation}
Integrating in time and volume and dividing by the total neutrino energy, $%
4\pi r^{2}n_{\nu }E_{\nu }\Delta t$, one obtains the fraction of total
energy transferred per neutrino, $\Delta E/E_{\nu }=\int \!dr\,\dot{\rho}%
/E_{\nu }n_{\nu }$.

For the sake of argument let us assume that some new interaction produces a
contribution like $A$ in Eq.~(\ref{Ahydrodynamic}), but without the factor $%
(1-\omega _{pl}^{2}/{\bf k}^{2})^{2}$, so that a resonance is possible for $%
{\bf k}$\ parallel to${\bf \ v}_{\nu }$ ($\omega _{pl}=k$) making $A\propto
\gamma ^{-1}$. The collisionless dispersion relation (\ref{dr})\ gives a
growth rate $\gamma \propto \Gamma ^{1/2}\omega _{pl}$ that is proportional
to $G_{{\rm F}}$ rather than $G_{{\rm F}}^{2}$. However, these resonant
modes are limited to the very thin shell $k=\omega _{pl}\pm \gamma $ in $%
{\bf k}$\ space. That makes an extra factor of $\gamma $ in the integration
of Eq.~(\ref{rhodot}) and the transferred energy goes as $\gamma ^{2}$,
proportional to $G_{{\rm F}}^{2}$ not to $G_{{\rm F}}$. Taking only the main
factors, one obtains $\Delta E/E_{\nu }\sim G_{{\rm F}}^{2}T_{e}\omega
_{pl}^{6}E_{\nu }^{-2}r$, about $10^{-18}$ for the same numbers used before
and $T_{e}=100\,{\rm KeV}$. This is very far from the few percent needed for
a Supernova explosion. It discourages the application of non-standard weak
interactions seeking for non-linear resonant effects.

\section{Conclusions}

We analyzed different instabilities that could possibly emerge from the
interaction between neutrinos and electron plasma waves in a Supernova
environment. The hypothetical neutrino induced resonances are too narrow to
embrace the total angular spread of the neutrino stream due to the finite
size of the neutrinosphere \cite{bento98,bento99}. But they are also too
narrow to contain the boundary regions of the neutrino phase space
distribution. When the resonant neutrino velocity vector is parallel to the
wavevector, a neutrino mass is needed and the energy spectrum is too broad
to keep the neutrino speed, $v_{\nu }\simeq 1-m_{\nu }^{2}/2E_{\nu }^{2}$,
inside the resonance. When the resonant velocities are in the boundary of
the velocity angular distribution, the depth of the neutrinosphere is too
large to make an angular boundary abrupt enough. In both cases the
boundaries of the neutrino phase space distribution are too wide to prevent
a departure from the poles to both sides of the resonances. As a result,
non-linear resonant effects are not possible and the growth or damping rates
are always linear in the neutrino flux and $G_{{\rm F}}^{2}$. They
correspond to a balance between stimulated \v{C}erenkov\ emission and
absorption of plasma waves by neutrinos \cite{hardy97}, analogous to Landau
damping. The energy that could possibly be transferred from neutrinos to
plasma waves seems to be vanishing small even for non-standard neutrino
interactions.

\section*{Acknowledgments}

This work was supported in part by FCT under grant PESO/P/PRO/1250/98.

\newpage

\section*{Appendix: Neutrinosphere depth}

Neutrinos like any other form of radiation are not emitted from an ideally
surface but from a shell with a certain depth. The neutrino 'optical' depth
at a point of radius $r$ can be defined as the Roseland depth, $\tau
(r)=\int_{r}^{\infty }dr/\lambda (r)$, where $\lambda ^{-1}=k\,\rho =\sigma
\,n$ is the inverse mean free path, a function of the opacity $k$ and
density $\rho $, or cross section $\sigma $ and number density of target
particles, $n$. $\tau (r)$ is a measurement of the number of collisions the
neutrinos suffer moving from the radius $r$ to infinity. The neutrinosphere
can be arbitrarily defined as the surface, of radius $R$, where $\tau =2/3$,
but the bulk of neutrino last scatterings spreads between the $\tau =1$ and $%
\tau =1/3$ surfaces. The way this translates as a radial distribution
depends on the chemical and density profile of the medium. For an
exponentially decreasing density near the neutrinosphere, $\rho (r)\propto
10^{-r/\ell }$, the radial separation between $\tau =1$ and $\tau =1/3$ is
approximately $2\Delta R\approx \ell /2$. Numerical simulations~\cite{wilson}
show that the length scale $\ell $\ varies from around $R/5$ to $R/20$ at
different stages of Supernova. Then, $\Delta R$ is about $R/20$ to$\ R/80$.

The other source of radial spread is the opacity or cross section dependence
on the neutrino energy~\cite{burrows90}, typically $k\propto E_{\nu }^{2}$.
At constant 'optical' depth $\tau $, the neutrinosphere radius increases
with the energy at a rate that we estimate as $\delta R/\delta E_{\nu
}=2\ell /E_{\nu }\ln 10$. Since the neutrino energy spectrum~\cite
{mayle87,burrows90} has a standard deviation $\Delta E_{\nu }\approx \bar{E}%
_{\nu }/2.5$ we obtain a radial deviation $\Delta R\approx \ell /3\approx
R/15-R/60$.\ The joint effect of statistical fluctuations and energy
spectrum is a neutrinosphere depth varying between $2\Delta R\approx 0.23R$
and $0.06R$. This implies a finite width in the boundary of the neutrino
angular aperture at large distances from the Supernova core as discussed in
the text.

\end{document}